\documentclass{PoS}

\title{A global 86-GHz VLBI survey of compact radio sources}

\ShortTitle{A global 86 GHz VLBI survey}

\author{\speaker{Sang-Sung Lee}\\
        Max-Planck-Institut f\"ur Radioastronomie\\
        E-mail: \email{sslee@mpifr-bonn.mpg.de}}

\author{Andrei P. Lobanov, Thomas P. Krichbaum, Arno Witzel, Anton Zensus\\
        Max-Planck-Institut f\"ur Radioastronomie\\
        E-mail: \email{alobanov@mpifr-bonn.mpg.de},  \email{tkrichbaum@mpifr-bonn.mpg.de},\\
                \email{awitzel@mpifr-bonn.mpg.de}, \email{azensus@mpifr-bonn.mpg.de} }

\author{Michael Bremer, Albert Greve, Michael Grewing \\
        Institut de Radio Astronomie Millim\'etrique\\
        E-mail: \email{bremer@iram.fr}, \email{greve@iram.fr}, \email{grewing@iram.fr} }

\abstract{
We present results from a large global VLBI\
survey of compact radio sources at 86 GHz\
begun in October 2001.\ 
The main goal of the survey was to increase \
the total number of objects\ 
accessible for future 3-mm VLBI imaging by factors of 3 $\sim$ 5.\  
The survey data attained the baseline sensitivity  of ~0.1\,Jy,\ 
and image sensitivity of better than 10 mJy/beam.\ 
To date, a total of 127 compact radio sources have been observed.\
The observations have yielded images for 109 sources, \
and only 6 sources have\
not been detected\footnote{
             The remaining 12 objects have been detected\
             but could not be imaged due to insufficient\
             closure phase information.}.
Flux densities and sizes of core and jet\ 
components of all
detected sources have been measured using Gaussian model fitting.\
From these measurements, brightness temperatures have been estimated, \
taking into account resolution limits of the data. \
Here, we compare the brightness temperatures of the cores and \
secondary jet components \
with similar estimates obtained from surveys at longer wavelengths\
 (e.g. 15\,GHz).\
This approach can be used to study the phenomena related to mechanisms \
of initial jet acceleration\
(accelerating or decelerating sub-pc jets?) and jet composition\
(electron-positron or electron-proton plasma?).}

\FullConference{The 8th European VLBI Network Symposium\\
		September 26-29, 2006\\
		Toru\'n, Poland}

\begin{document}

\section{Observations}

      Millimetre-wavelength VLBI (mm-VLBI) provides a unique tool for exploring
      the physical nature of compact radio sources. 
      The higher resolution\footnote{Global VLBI observations 
                            at 86\,GHz offer about 6 times better
                            resolution than space VLBI observations at 5\,GHz.}
      of VLBI observations at millimetre wavelengths
      allows us to image directly the VLBI core and the knots 
      with subparsec-scale resolution. Furthermore, in AGN, synchrotron
      radiation becomes optically
      thin at wavelengths between 1\,cm and 1\,mm \cite{Stevens94}. Therefore, 
      the mm-VLBI imaging makes it possible to look deeper into the VLBI cores
      that are invisible at centimetre wavelengths.
      The first detection of single-baseline interference fringes of 89\,GHz (3.4\,mm) 
      VLBI observation
      was reported by \cite{Readhead83}, demonstrating the feasibility of 3-mm VLBI.
      A decade later, the first global mm-VLBI array, 
      the Coordinated Millimeter VLBI Array (CMVA),
      was established in 1995 with 
      support from radio observatories throughout the world.
      The number of participating telescopes gradually increased up to 12.
      The CMVA discontinued to organize mm-VLBI experiments in 2002.
      Since then, the activity of mm-VLBI experiments has been continued through 
      the Global mm-VLBI Array (GMVA).
      The GMVA carries out regular, coordinated global VLBI observations 
      at 86\,GHz, providing good quality images with a typical angular resolution 
      of 50-70 micro-arcseconds ($\mu as$).  
      In order to increase the number of objects imaged at 86\,GHz,
      four detection and imaging surveys have been conducted to date, with a total of
      124 extragalactic radio sources observed at 86\,GHz
      (see \cite{Beasley96}, \cite{Lonsdale98}, \cite{Ranta98}, \cite{Lobanov00}).
      Fringes have been detected in 44 objects, but only 24 radio sources have been
      successfully imaged so far.
      Table\,\ref{table:vlbisurveys} gives an overview of the VLBI surveys at 86\,GHz.
      The survey presented here was envisaged as a project that would increase
      the number of objects imaged at 86\,GHz by a factor of 3 to 5.

       The source selection of this survey was based on the results 
       from the VLBI surveys at 22\,GHz\,\cite{Moellen96},
       15 GHz\,\cite{Kellermann98}, and on source fluxes 
       obtained from the multifrequency 
       monitoring data from Mets\"ahovi at 22, 37,
       and 86\,GHz\,\cite{Teras98} and from Pico Veleta 
       at 90, 150, and 230\,GHz (Ungerechts, priv. comm.).
       Using these databases, we selected the sources with
       expected flux density above $\geq 0.3$\,Jy at 86\,GHz. 
       We excluded some of the brightest sources 
       already imaged at 86\,GHz,
       and focused on those sources that had not 
       been detected or imaged in the previous surveys.
       In order to optimize the {\it uv}-coverage
       of the survey data,
       sources at higher northern declinations ($\delta > -40^{\circ}$)
       were preferred.
       According to the aforementioned selection criteria, 
       a total of 127 compact radio sources were selected
       and observed. 

      The survey observations were conducted during 
      three global mm-VLBI sessions on October 2001,
      April 2002 and October 2002.
      The participation
      of the large and sensitive European antennas 
      (the 100-m radio telescope at Effelsberg, the
      30-m mm-radio telescope at Pico Veleta, the 6$\times$15-m interferometer 
      on Plateau de Bure) resulted in a typical 
      single baseline sensitivity of $\sim 0.1$\,Jy and
      an image sensitivity of better than 10\,mJy/beam.
      Every source in the survey was observed for 3-4 scans 
      of 7-min duration ({\it snapshot} mode).
      The data were recorded either with 128-MHz 
      or 64-MHz bandwidth
      using the MkIV VLBI system with 1- and 2- bit sampling
      adopted at different epochs.
      The observations were made in left circular polarization (LCP).
      3 to 4 scans per hour were recorded, using the time between 
      the scans for antenna focusing, pointing
      and calibration.
      The data were correlated using the MkIV 
      correlator of Max-Planck-Institut f\"ur
      Radioastronomie (MPIfR) in Bonn. 
      Fringes were searched with the HOPS package 
      {\it fourfit} and AIPS%
      \footnote{The NRAO Astronomical Image Processing System.}
      task
      {\bf} FRING.
      The amplitude calibration was made 
      using the system temperature,  
      antenna gain, and opacity measurements 
      carried out at each station 
      during the observations. 
      The AIPS task APCAL was used to calibrate the amplitudes.
      From the phase- and amplitude-calibrated data,
      the images were made using DIFMAP software \cite{Shepherd94}.
      The detailed description of the fringe-fitting 
      and imaging will be given in \cite{Lee2006}.

\begin{table}[]
  \caption{{\footnotesize VLBI surveys at 86 GHz}}
\begin{minipage}{1\textwidth}
\begin{center}
\bigskip
\normalsize
\begin{tabular}{cccccccc}\hline\hline
Surveys & $N_{\rm ant}$  & $\Delta S$  &  $\Delta I_{\rm m}$ &  $D_{\rm img}$ & $N_{\rm obs}$  & $N_{\rm det}$ & $N_{\rm img}$ \\
(1)  & (2)        & (3)        &  (4)         &  (5)       & (6)        & (7)       & (8)       \\\hline
1    & 3          & $\sim$0.5  &  ...         &  ...       & 45         & 12        & ...       \\
2    & 2$-$5      & $\sim$0.7  &  ...         &  ...       & 79         & 14        & ...       \\
3    & 6$-$9      & $\sim$0.5  &  $\sim$30    &  70        & 67         & 16        & 12        \\
4    & 3$-$5      & $\sim$0.4  &  $\sim$20    &  100       & 28         & 26        & 17        \\\hline
\multicolumn{5}{c}{Total number of unique objects:}        & 124        & 44        & 24        \\\hline
\multicolumn{8}{l}{Properties of this survey:}\\
     & 12$-$14    & $\sim$0.2 &   $\leq$10    &  50        & 127        & 121       & 109       \\\hline
\hline
\end{tabular}
\label{table:vlbisurveys}%
\end{center}
\noindent
\\
{\footnotesize Notes: {\bf Columns:} 1 -- references of previous VLBI surveys at 86\,GHz; 
       2 -- number of participating antennae; 3 -- average baseline sensitivity [Jy];
       4 -- average image sensitivity [mJy/beam]; 5 -- typical dynamic range of images; 
       6 -- number of sources observed;
       7 -- number of objects detected; 8 -- number of objects imaged. 
                          {\bf References of surveys:}
                          1 -- \cite{Beasley96}; 
                          2 -- \cite{Lonsdale98}; 
                          3 -- \cite{Ranta98}; 
                          4 -- \cite{Lobanov00}.
}
\end{minipage}
\end{table}

\section{Images}

\begin{figure}
\includegraphics[width=1.0\textwidth]{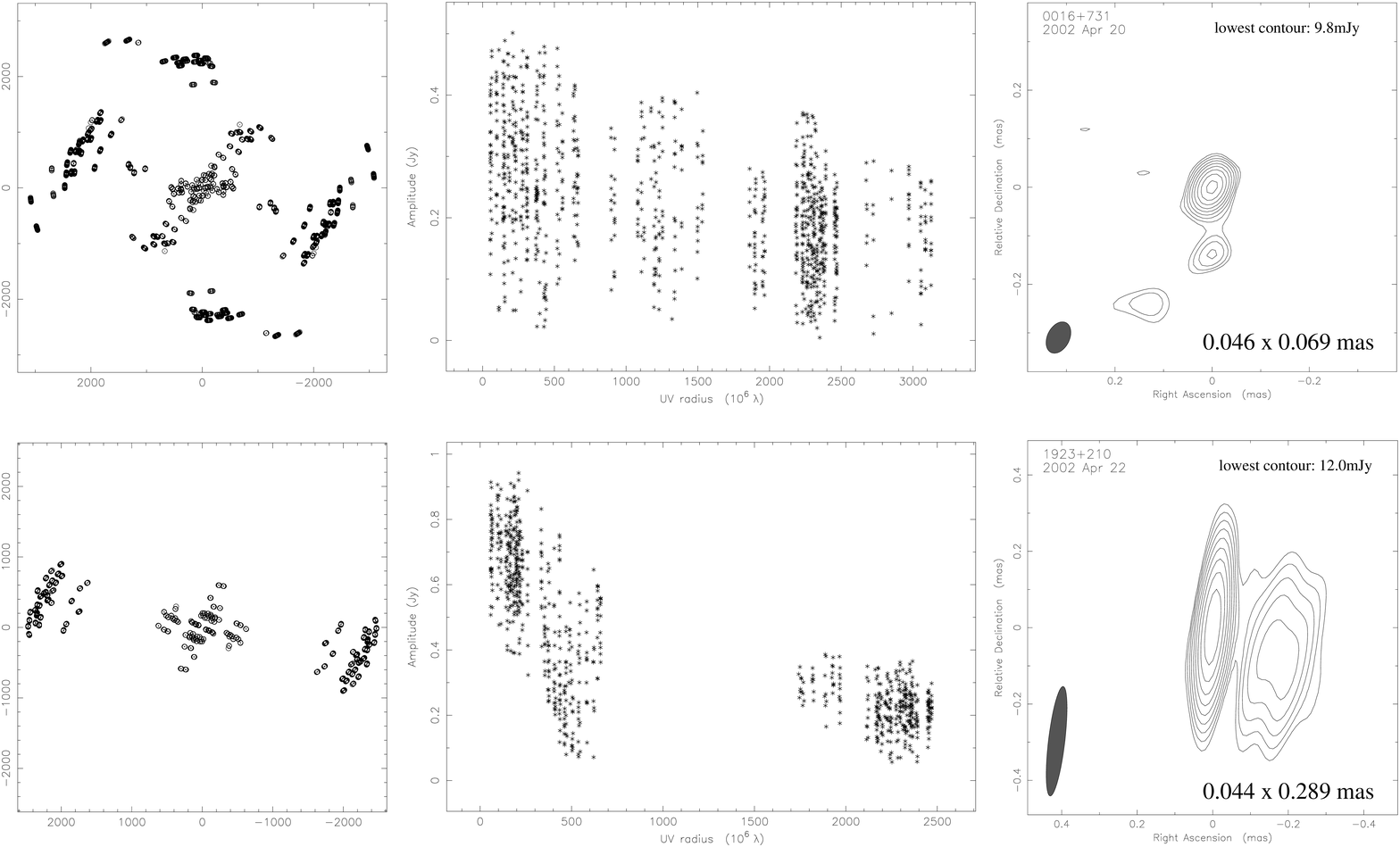}
\caption{{\footnotesize Images with the distribution of the {\it uv}-sampling 
         and the visibility amplitudes 
         of 0016+731 and 1923+210.
         The contours are drawn at -1,1,1.4,...,$1.4^n$ of 
         the lowest flux density levels, 
         9.8\,mJy and 12.0\,mJy, respectively.}}
\label{fig:map}
\end{figure}

      We have detected 121 sources and have produced hybrid maps 
      of 109 sources for which the data contain
      a sufficiently large number of {\it uv}-points. 
      Out of 109 sources, 
      90 sources are imaged for the first time at 86\,GHz,
      increasing the number of sources ever imaged with 
      86-GHz VLBI observations up to 110. To illustrate our results,
      we present here the images for two relatively weak objects.
      These are the first 3-mm VLBI maps 
      for the respective sources. 
      For each image, plots of the {\it uv}-coverage  
      and of the visibility amplitudes against
      {\it uv}-radius are presented. 
      Here we report details on two sources.

      \noindent
      {\bf 0016+731:} \\
      The image in Figure \ref{fig:map} (top, right) shows two jet components 
      along the different directions,
      $P.A. \sim -179^{\circ}$ and $P.A. \sim 150^{\circ}$, 
      appearing to be bent to the southeast. 
      The direction of the jet is in a good
      agreement with the VLBI images of 0016+731 obtained 
      at 43\,GHz\,\cite{Lister01} 
      and 15\,GHz\,\cite{Lister05}. The peak flux density
      is 0.196\,Jy/beam with a beam size of $0.046 \times 0.069$\,mas 
      and the lowest contour level is 9.8\,mJy.
      The brightness temperature ($T_{\rm b}$) of the core component 
      is ($2.5\pm0.8)\times 10^{11}$\,K. 
      Since the source
      is at high declination, the {\it uv}-sampling (shown in the left 
      panel) is nearly circular resulting in a very regular restoring beam.
      A decrease of the visibility amplitude
      with the {\it uv}-distance (middle panel) indicates that
      the source is resolved.

      \noindent
      {\bf 1923+210:} \\
      In the image in Figure \ref{fig:map} (bottom, right), we identify one feature 
      in the jet extending along 
      $P.A. \sim -114^{\circ}$, which is similar to the orientation
      of the jet observed at lower frequencies\cite{Fey00}. 
      The peak flux density is 0.301\,Jy/beam, with a beam size
      of $0.044 \times 0.289$\,mas, and the lowest contour level is 12.0\,mJy.
      The brightness temperature of the core component is ($5.1\pm1.8)\times 10^{10}$\,K.

\section{Brightness temperature and jet physics}

\begin{figure}
\includegraphics[width=1.0\textwidth]{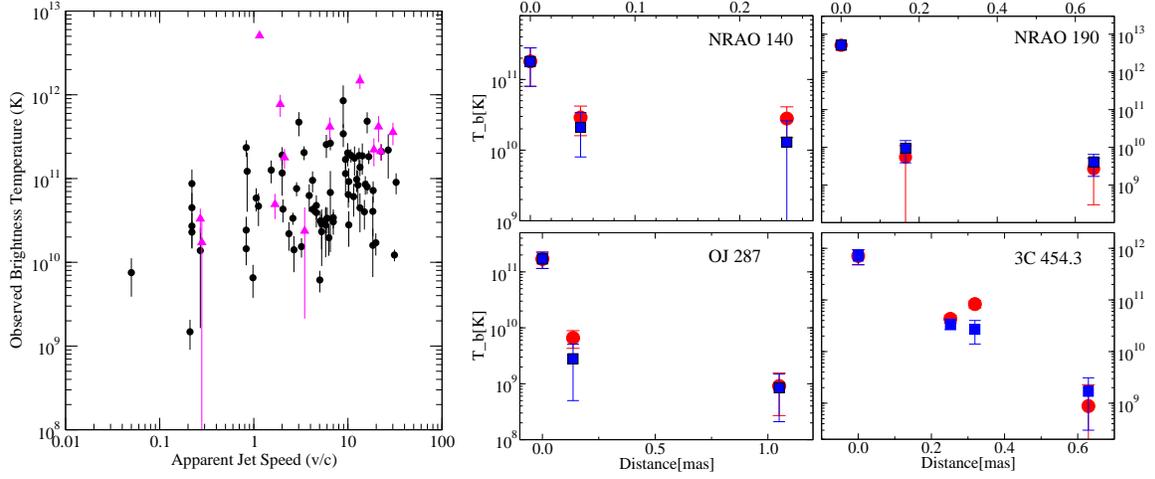}
\caption{{\bf Left:} A plot of measured $T_{\rm b}$ in the cores of the jets vs. maximum apparent jet speeds
          taken from  2-cm survey. Triangles are the lower limits.
         {\bf Right:} Changes of the brightness temperature along the jets of several sources.
         Blue squares are the measured $T_{\rm b}$. Red circles are the predicted $T_{\rm b}$ 
         in adiabatically expanding shocks with the initial brightness temperature
         equal to that measured in the core of the jet.
         }
\label{fig2}
\end{figure}

     Using the Gaussian-model fitted images of 109 sources, we have 
     determined basic properties of the core and jet components
     for all of the imaged sources: total flux density, $S_{\rm tot}$,
     peak flux density, $S_{\rm peak}$, post-fit rms, $\sigma_{\rm rms}$, size, $d$,
     radial distance, $r$ (for jet components), 
     and position angle, $\theta$ (for jet components). 
     For all parameters,
     the uncertainties of measurements have been obtained, taking into account
     the resolution limits\,\cite{Lob2005}: \\
     \begin{equation}
          d_{\rm min}=\frac{2^{1+\beta/2}}{\pi}{\left[{\pi ab\ln2\ln{\frac{SNR}{SNR-1}}}\right]}^{1/2},  \\
     \end{equation}

     \noindent
     which is the minimum resolvable size of a component in an image. 
     The minimum resolvable size depends on the axes of the restoring beam, {\it a} and {\it b}, and 
     the signal-to-noise ratio, {\it SNR}. The weighting function, $\beta$, is 0 for natural weighting
     or 2 for uniform weighting. The observed brightness temperature, $T_{\rm b}$, of a component in the 
     rest frame of the source is given by:\\

     \begin{equation}
      T_{\rm b} = \frac{2\ln{2}}{{\pi}{k_{\rm B}}}\frac{S_{\rm tot}\lambda^2(1+z)}{d^2}, \\
     \end{equation}

     \noindent
     where $\lambda$ is the wavelength of observation, $z$ is the redshift, and $k_{\rm B}$ is the 
     Boltzmann constant.
     If $d < d_{\rm min}$, then the lower limit of $T_{\rm b}$ is obtained 
     with $d=d_{\rm min}$. The observed brightness temperatures can be used to study the physics of 
     the relativistic jets. One of the applications is to model the observed distribution of $T_{\rm b}$ 
     by a population of jets in which all jets are randomly oriented and 
     have the same Lorentz factor, $\gamma_{\rm j}$, the same
     spectral index, $\alpha$, and the same intrinsic brightness temperature, $T_{\rm 0}$ (see \cite{Lobanov00}).
     Since the orientation of the jets is random, and the observed distribution of $T_{\rm b}$ is caused
     only by Doppler boosting, the predicted distribution should be corrected for the bias due to 
     the Doppler boosting. Lobanov et al.\,\cite{Lobanov00} applied this approach to a smaller sample
     of VLBI images at 86 GHz, inferring the range of $T_{\rm 0} \sim 1-4 \times 10^{11}$\,K that reproduces
     the observed distribution in the VLBI cores. We expect a similar result for our larger sample.
     Another application is to study the intrinsic brightness
     temperatures of the VLBI cores, by using the observed $T_{\rm b}$ at 86 GHz from our survey and 
     the maximum apparent jet speeds at 15 GHz taken from \cite{Kellermann98}.
     In order to constrain the intrinsic brightness temperature,
     the method from \cite{Homan2006} could also be applied to our sample. 
     In the left panel of Figure\,\ref{fig2}, 
     we can see a hint of correlation of the observed brightness temperatures of the VLBI cores
     with the maximum apparent jet speeds of the jets measured at 15 GHz.
     The right panel of Figure\,\ref{fig2} shows the profile of $T_{\rm b}$ along the main jet direction.
     The values of brightness temperatures in the jets predicted assuming adiabatic expansion
     in a relativistic plasma\,\cite{Lobanov00} agree well
     with the observed brightness temperatures.
     Using these three approaches to analyze the 86-GHz data,
     we hope to be able to further constrain physical models proposed to explain 
     the nature of compact relativistic jets.\\\\

\noindent
{\bf Acknowledgements}\\

We gratefully thank the staff of the observatories participating in the GMVA;  
the MPIfR Effelsberg 100-m telescope, the two IRAM telescopes on Plateau de Bure and
Pico Veleta, the Mets\"ahovi Radio Observatory, the Onsala Space Observatory, and the VLBA.
The VLBA is an instrument of the National Radio Astronomy Observatory, which is
a facility of the National Science Foundation operated under cooperative agreement 
by Associated Universities, Inc.


\begin{thebibliography}{99}
\bibitem{Stevens94} 
 J. A. Stevens, S. J. Litchfield, E. I. Robson, D. H. Hughes, W. K. Gear, 
 H. Ter\"asranta, E. Valtaoja, and M. Tornikoski, 
 \emph{Multifrequency observations of blazars. 5: Long-term millimeter, submillimeter, and 
        infrared monitoring},
 \emph{ApJ} {\bf 437} (1994) 91

\bibitem{Readhead83} 
 A. C. S. Readhead, C. R. Mason, A. T. Moffet, et al.,
 \emph{Very long baseline interferometry at a wavelength of 3.4\,mm},
 \emph{Nature} {\bf 303} (1983) 504

\bibitem{Beasley96} 
 A. J. Beasley, V. Dhawan, S. Doeleman, and R. B. Phillips,
 \emph{CMVA Observations of Compact AGNs},
 \emph{Proceedings of Millimeter-VLBI Science Workshop} (1997) 53 

\bibitem{Lonsdale98} 
 C. J. Lonsdale, S. S. Doeleman, and R. B. Phillips,
 \emph{A 3 Millimeter VLBI Continuum Source Survey},
 \emph{AJ} {\bf 116} (1998) 8 

\bibitem{Ranta98} 
 F. T. Rantakyr\"o, L. B. Baath, D. C. Backer, et al.,
 \emph{50 MU as resolution VLBI images of AGN's at lambda 3 MM},
 \emph{A\&AS} {\bf 131} (1998) 451 

\bibitem{Lobanov00} 
 A. P. Lobanov, T. P. Krichbaum, D. A. Graham, et al.,
 \emph{86 GHz VLBI survey of compact radio sources},
 \emph{A\&A} {\bf 364} (2000) 391 

\bibitem{Moellen96} 
 G. A. Moellenbrock, K. Fujisawa, R. A. Preston, et al., 
 \emph{A 22 GHz VLBI Survey of 140 Compact Extragalactic Radio Sources},
 \emph{AJ} {\bf 111} (1996) 2174 

\bibitem{Kellermann98} 
 K. I. Kellermann, R. C. Vermeulen, J. A. Zensus, and M. H. Cohen,
 \emph{Sub-Milliarcsecond Imaging of Quasars and Active Galactic Nuclei},
 \emph{AJ} {\bf 115} (1998) 1295 

\bibitem{Teras98} 
 H. Ter{\"a}sranta, M. Tornikoski, A. Mujunen, et al.,
 \emph{Fifteen years monitoring of extragalactic radio sources at 22, 37 and 87 GHz},
 \emph{A\&AS} {\bf 132} (1998) 305 


\bibitem{Shepherd94} 
 M. C. Shepherd, T. J. Pearson, and G. B. Taylor,
 \emph{Software Report: DIFMAP, Owens Valley Radio Observatory},
 \emph{BAAS} {\bf 26} (1994) 987 

\bibitem{Lee2006} 
 Sang-Sung Lee,
 \emph{A Global 86 GHz VLBI Survey of Compact Radio Sources},
 \emph{PhD Thesis} at University of Bonn, in prep. 

\bibitem{Lister01} 
 M. Lister,
 \emph{Parsec-Scale Jet Polarization Properties of a Complete Sample of Active Galactic Nuclei at 43 GHz},
 \emph{ApJ} {\bf 562} (2001) 208. 

\bibitem{Lister05} 
 M. Lister and D. C. Homan,
 \emph{MOJAVE: Monitoring of Jets in Active Galactic Nuclei with VLBA Experiments. I. First-Epoch 15 GHz Linear Polarization Images},
 \emph{AJ} {\bf 130} (2005) 1389. 

\bibitem{Fey00} 
 A. L. Fey and P. Charlot,
 \emph{VLBA Observations of Radio Reference Frame Sources. III. Astrometric Suitability of an Additional 225 Sources},
 \emph{ApJS} {\bf 128} (2000) 17. 

\bibitem{Lob2005} 
 A. P. Lobanov,
 \emph{Resolution limits in astronomical images},
 \emph{A\&A}, submitted (astro-ph/0503225).

\bibitem{Homan2006} 
 D. C. Homan, Y. Y. Kovalev, M. L. Lister, et al.
 \emph{Intrinsic Brightness Temperatures of AGN Jets},
 \emph{ApJ}, {\bf 642} L115.



\end{thebibliography}
\end{document}